\documentclass[aps,prb,fleqn,12pt,superscriptaddress]{revtex4}
\usepackage{epsfig,color}
\usepackage{graphicx}
\usepackage{pdfpages}
\usepackage{amssymb,amsmath}
\usepackage{hyperref}

\begin{document}
\title{Spin-orbit coupling induced magnetic anisotropy \\ and large spin wave gap in {\boldmath $\rm Na Os O_3$}}
\author{Avinash Singh}
\email{avinas@iitk.ac.in}
\affiliation{Department of Physics, Indian Institute of Technology, Kanpur - 208016, India}
\affiliation{Department of Physics and Astronomy, University of Missouri, Columbia, MO 65211}
\author{Shubhajyoti Mohapatra}
\affiliation{Department of Physics, Indian Institute of Technology, Kanpur - 208016, India}
\author{Churna Bhandari}
\affiliation{Department of Physics and Astronomy, University of Missouri, Columbia, MO 65211} 
\author{Sashi Satpathy}
\affiliation{Department of Physics and Astronomy, University of Missouri, Columbia, MO 65211} 
\date{\today} 


\begin{abstract}
The role of spin-orbit coupling and Hund's rule coupling on magnetic ordering, anisotropy, and excitations are investigated within a minimal three-orbital model for the $5d^3$ compound $\rm Na Os O_3$. Small asymmetry between the magnetic moments for the $xy$ and $xz,yz$ orbitals, arising from the hopping asymmetry generated by $\rm Os O_6$ octahedral tilting and rotation, together with the weak correlation effect, are shown to be crucial for the large SOC induced magnetic anisotropy and spin wave gap observed in this compound. Due to the intrinsic SOC-induced changes in the electronic densities under rotation of the staggered field, their coupling with the orbital energy offset is also found to contribute significantly to the magnetic anisotropy energy.

\end{abstract}
\pacs{75.30.Ds, 71.27.+a, 75.10.Lp, 71.10.Fd}
\maketitle
\newpage

\section{Introduction}

The strongly spin-orbit coupled orthorhomic structured $5d^3$ osmium compound $\rm Na Os O_3$, with nominally three electrons in the Os $t_{2g}$ sector, exhibits several novel electronic and magnetic properties. These include a G-type antiferromagnetic (AFM) structure with spins oriented along the $c$ axis,\cite{calder_PRL_2012} a significantly reduced magnetic moment $\sim 1 \mu_{\rm B}$ as measured from neutron scattering,\cite{calder_PRL_2012} a continuous metal-insulator transition (MIT) that coincides with the AFM transition ($T_{\rm N} = T_{\rm MIT}$ = 410 K) as seen in neutron and X-ray scattering,\cite{calder_PRL_2012} and a large spin wave gap of 58 meV as seen in resonant inelastic X-ray scattering (RIXS) measurements indicating strong magnetic anisotropy.\cite{calder_PRB_2017} 

Neutron scattering and RIXS studies of the magnetic excitation spectrum have also revealed large spin wave gap in the frustrated type I AFM ground state of the double perovskites $\rm Ba_2 Y Os O_6$, $\rm Sr_2 Sc Os O_6$, $\rm Ca_3 Li Os O_6$,\cite{kermarrec_PRB_2015,taylor_PRB_2016,taylor_PRL_2017} highlighting the importance of SOC-induced magnetic anisotropy despite the nominally orbitally-quenched ions in the $5d^3$ and $4d^3$ systems. For the pyrochlore compound $\rm Cd_2 Os_2 O_7$ also, neutron diffraction and RIXS measurements have directly probed the $5d$ electrons responsible for the magnetic order and MIT in both the metallic and insulating regimes.\cite{calder_Nature_2016}

Investigations of the electronic and magnetic properties using first-principles calculations have been carried out for the orthorhombic perovskite $\rm Na Os O_3$,\cite{du_PRB_2012,jung_PRB_2013} related osmium based perovskites $\rm AOsO_3$ (A=Ca,Sr,Ba),\cite{zahid_JPCS_2015} and double perovskites $\rm Ca_2 Co Os O_6$ and $\rm Ca_2 Ni Os O_6$.\cite{morrow_CM_2016} Density functional theory (DFT) calculations have shown that the magnetic moment is strongly reduced to nearly $1 \mu_{\rm B}$ (essentially unchanged by SOC) due to itineracy resulting from the strong hybridization of Os $5d$ orbitals with O $2p$ orbitals, which is significantly affected by the structural distortion.\cite{jung_PRB_2013} Furthermore, from total energy calculations for different spin orientations with SOC included, the easy axis was determined as $\langle 001 \rangle$,\cite{jung_PRB_2013} as also observed by Calder {\em et al.},\cite{calder_PRL_2012} with large energy cost for orientation along the $\langle 010 \rangle$ axis and very small energy difference between orientations along the nearly symmetrical $a$ and $c$ axes. 


Although weak correlation effects are central to the electronic and magnetic behavior for both $\rm Na Os O_3$ and $\rm Cd_2 Os_2 O_7$ which exhibit continuous MIT concomitant with three dimensional AFM ordering, magnetic interactions and excitations in both compounds have been studied only within the phenomenological localized spin picture. Investigation of the strong SOC-induced magnetic anisotropy energy (MAE) and spin wave gap within the itinerant electron picture in terms of a weakly correlated minimal three-orbital model is therefore of particular interest. For the iridate compounds, recent study of magnetic excitations in terms of the itinerant electron approach has provided a microscopic understanding of features such as the strong zone boundary spin wave dispersion in the single-layer compound and the large spin wave gap in the bilayer compound, as observed in RIXS studies, in terms of characteristic weak correlation effects in the $5d$ systems.\cite{shubhajyoti_PRB_2017}


In this paper, we will therefore investigate: i) the key features required in a minimal three-orbital model within the $t_{2g}$ sector in order to understand the SOC-induced magnetic anisotropy and preferred ordering direction, ii) role of the Hund's coupling term on the magnetic order, and (iii) magnetic excitations and the large spin wave gap. The Hund's coupling term has a particularly important role in view of the SOC-induced intra-site magnetic frustration (similar to that in the triangular-lattice AFM) due to Kitaev type anisotropic spin interactions involving the magnetic moments $S_\mu$ for the three orbitals. 

The structure of this paper is as follows. Starting with a minimal three-orbital model in Sec. II, the SOC-induced magnetic anisotropy in the AFM state is studied in Sec. III for different orientations of the staggered field. Here the staggered fields (and therefore the magnetic moments $m_\mu$) for the three orbitals are assumed to be parallel. As this orbitally collinear AFM state is not the ground state in the absence of Hund's coupling, an orbitally canted AFM state is studied in Sec. IV, motivated by the SOC-induced anisotropic spin interactions and intra-site magnetic frustration effect (Appendix). Magnetic excitations are studied in Sec. V, highlighting the non-trivial role of Hund's coupling in overcoming the magnetic frustration, stabilizing the orbitally collinear AFM state, and activating the SOC-induced magnetic anisotropy. Finally, the role of orbital energy offset on MAE is investigated in Sec. VI, and conclusions are presented in Sec. VII.

\section{Three orbital model and magnetic ordering}

A complex interplay between SOC, structural distortion, magnetic ordering, Hund's rule coupling, and weak correlation effect is evident from the electronic and magnetic behaviour of $\rm Na Os O_3$ as discussed above. While strong Hund's rule coupling ($J_{\rm H}$) would favor high-spin $S=3/2$ state in the half-filled system with three electrons per Os ion, spin-orbital entangled states energetically separated into the $J=1/2$ doublet and $J=3/2$ quartet would be favored by strong SOC. In the formation of the AFM state, the weak correlation term is supported by $J_{\rm H}$ which effectively enhances the local exchange field, thus also self consistently suppressing the SOC by  energetically separating the spin up and down states. 

A detailed study of the electronic band structure of $\rm Na Os O_3$ has been carried out recently for both the undistorted and distorted structures.\cite{osmate_band_structure_2017} Effects of the structural distortion associated with the $\rm OsO_6$ octahedral rotation and tilting on the electronic band structure were investigated using the density functional theory (DFT) and reproduced within a realistic three-orbital model. The orbital mixing terms resulting from the octahedral rotations were shown to account for the fine features in the DFT band structure. Study of staggered magnetization indicated weak coupling behavior, and the small moment disparity ($m_{yz},m_{xz} > m_{xy}$) obtained for the distorted structure reflected a relative bandwidth reduction for the $yz,xz$ orbitals.
 

In order to investigate the SOC induced magnetic anisotropy and large spin wave gap in this AFM insulating system, we will consider a minimal three-orbital model involving the $yz,xz,xy$ orbitals within the $t_{2g}$ sector at half filling ($n=3$). The role of the structural distortion will be incorporated through a small hopping (bandwidth) asymmetry broadly consistent with the electronic band structure comparison mentioned above.

Combining the SOC, band, and staggered field terms, the Hamiltonian in the composite three-orbital $(yz\sigma,xz\sigma,xy\bar{\sigma})$, two-sublattice ($s=\pm 1$) basis is obtained as:\cite{osmate_band_structure_2017}
\begin{eqnarray}
\mathcal{H}_{\rm SO} + \mathcal{H}_{\rm band} + \mathcal{H}_{\rm sf} 
& = & \sum_{{\bf k} \sigma s} \psi_{{\bf k} \sigma s}^{\dagger} \left [ \begin{pmatrix}
{\epsilon_{\bf k} ^{yz}}^\prime & i \sigma\frac{\lambda}{2} & -\sigma\frac{\lambda}{2} \\
- i \sigma\frac{\lambda}{2} & {\epsilon_{\bf k} ^{xz}}^\prime & i\frac{\lambda}{2} \\
-\sigma\frac{\lambda}{2} & - i\frac{\lambda}{2} & {\epsilon_{\bf k} ^{xy}}^\prime \end{pmatrix} 
-s\sigma \begin{pmatrix}
\Delta_{yz}^z & 0 & 0 \\
0 & \Delta_{xz}^z & 0 \\
0 & 0 & -\Delta_{xy}^z \\
\end{pmatrix} \right ]
\psi_{{\bf k} \sigma s} \nonumber \\
& + & 
\sum_{{\bf k} \sigma s} \psi_{{\bf k} \sigma s}^{\dagger}
\begin{pmatrix}
\epsilon_{\bf k} ^{yz} & \epsilon_{\bf k} ^{yz|xz} & \epsilon_{\bf k} ^{yz|xy} \\
-\epsilon_{\bf k} ^{yz|xz} & \epsilon_{\bf k} ^{xz} & \epsilon_{\bf k} ^{xz|xy} \\
-\epsilon_{\bf k} ^{yz|xy} & -\epsilon_{\bf k} ^{xz|xy} & \epsilon_{\bf k} ^{xy} \end{pmatrix} 
\psi_{{\bf k} \sigma \bar{s}} 
\label{three_orb_two_sub}
\end{eqnarray}
which is defined with respect to a common spin-orbital coordinate system. Here $\lambda$ is the SOC constant, $\epsilon_{\bf k} ^{\mu}$ and ${\epsilon_{\bf k} ^{\mu}}^\prime$ are the band energies for the three orbitals $\mu$ corresponding to the hopping terms connecting same and opposite sublattices, respectively. Also included are the orbital mixing hopping terms $\epsilon_{\bf k} ^{\mu |\nu}$ arising from the octahedral rotation and tilting. All nearest-neighbor hopping terms are placed in the sublattice-off-diagonal $(s\bar{s})$ part of the Hamiltonian. The symmetry-breaking staggered field term is shown here for $z$ direction ordering. For general ordering direction with components {\boldmath $\Delta_\mu$}= $\Delta_\mu^x,\Delta_\mu^y,\Delta_\mu^z$, the staggered field term:  
\begin{equation}
\mathcal{H}_{\rm sf} = 
\sum_{{\bf k} \sigma \sigma' s \mu}  \psi_{{\bf k} \sigma s \mu}^{\dagger} 
\begin{pmatrix} -s \makebox{\boldmath $\sigma . \Delta_\mu$}
\end{pmatrix}_{\sigma \sigma'} \psi_{{\bf k} \sigma' s \mu} 
= \sum_{{\bf k} \sigma \sigma' s \mu} s \psi_{{\bf k} \sigma s \mu}^{\dagger} 
\begin{pmatrix} -\Delta_\mu ^z & -\Delta_\mu ^x + i \Delta_\mu ^y \\
-\Delta_\mu ^x - i \Delta_\mu ^y & \Delta_\mu ^z  \\
\end{pmatrix}_{\sigma \sigma'} \psi_{{\bf k} \sigma' s \mu}
\label{gen_ord_dirn} 
\end{equation} 
The staggered fields {\boldmath $\Delta_\mu$} are self-consistently determined from:
\begin{equation}
2\makebox{\boldmath $\Delta_\mu$} = U_\mu \makebox{\boldmath $m_\mu$} + 
J_{\rm H} \sum_{\nu \neq \mu} \makebox{\boldmath $m_\nu$}
\label{selfcon}
\end{equation}
in terms of the staggered magnetizations {\boldmath $m_\mu$}=($m_\mu ^x,m_\mu ^y,m_\mu ^z$) for the three orbitals $\mu$. The staggered field terms in the AFM state arise from the Hartree-Fock (HF) approximation of the electron interaction terms: $\sum_{i\mu} U_\mu n_{i\mu\uparrow} n_{i\mu\downarrow} - 2J_{\rm H} \sum_{i,\mu \ne \nu} {\bf S}_{i\mu} . {\bf S}_{i\nu}$, where $U$ and $J_{\rm H}$ are the Hubbard and Hund's rule coupling terms, respectively. For general ordering direction, the staggered magnetization components ($\alpha=x,y,z$) are evaluated from:
\begin{equation}
[m_\mu ^\alpha ]_A = \frac{1}{N} \sum_{{\bf k},l} ^{E_{{\bf k}l} < E_{\rm F}} 
\left ( \phi_{{\bf k}l\mu}^{\uparrow *} \; \; \phi_{{\bf k}l\mu}^{\downarrow *} \right )_A 
\; [\sigma ^\alpha ] \; \left ( \begin{array}{c}
\phi_{{\bf k}l\mu}^{\uparrow} \\ \phi_{{\bf k}l\mu}^{\downarrow}
\end{array} \right )_A = - [m_\mu ^\alpha ]_B
\label{magneqn}
\end{equation}
where $\phi_{{\bf k}l}$ are the eigenvectors of the Hamiltonian $\mathcal{H}_{\rm SO} + \mathcal{H}_{\rm band} + \mathcal{H}_{\rm sf}$, $l$ is the branch label and $N$ is the total number of ${\bf k}$ states. In practice, it is easier to consider a given $\Delta$ and self-consistently determine the interaction strength $U_\mu$ from Eq. \ref{selfcon}. 


Corresponding to the hopping terms in the tight-binding representation, we will consider the band energy contributions in Eq. (\ref{three_orb_two_sub}) for opposite ($\epsilon_{\bf k} ^{\mu}$) and same (${\epsilon_{\bf k} ^{\mu}}^\prime$) sublattices: 
\begin{eqnarray}
\varepsilon^{xy}_{\bf k} &=& -4t_1 \cos{(k_x/2)} \cos{(k_y/2)} \nonumber \\
{\varepsilon^{xy}_{\bf k}}^{\prime} &=& - 2t_2 (\cos{k_x} + \cos{k_y}) + \epsilon_{xy}  \nonumber \\
\varepsilon^{yz}_{\bf k} &=& -2t_4 [ \cos{\{(k_x - k_y)/2\}} + \cos{k_z}] \nonumber \\
{\varepsilon^{yz}_{\bf k}}^{\prime} &=& -4 t_2 \cos{\{(k_x - k_y)/2\}} \cos{k_z} \nonumber \\
\varepsilon^{xz}_{\bf k} &=& -2t_4 [\cos{\{(k_x + k_y)/2\}} + \cos{k_z}] \nonumber \\
{\varepsilon^{xz}_{\bf k}}^{\prime} &=& -4 t_2 \cos{\{(k_x + k_y)/2\}} \cos{k_z} \nonumber \\
\varepsilon^{yz|xz}_{\bf k} &=& -4t_{m1} \cos{(k_x/2)} \cos{(k_y/2)}  \nonumber \\
\varepsilon^{yz|xy}_{\bf k} &=& +2t_{m2} [ \cos{\{(k_x + k_y)/2\}} + 2\cos{\{(k_x - k_y)/2\}} + \cos{k_z} ] \nonumber \\
\varepsilon^{xz|xy}_{\bf k} &=& -2t_{m2} [ 2\cos{\{(k_x + k_y)/2\}} + \cos{\{(k_x - k_y)/2\}} + \cos{k_z} ] 
\label{three_orb_model}
\end{eqnarray}
Here $t_1$ and $t_2$ are the first and second neighbor hopping terms for the $xy$ orbital, which has energy offset $\epsilon_{xy}$ relative to the degenerate $yz/xz$ orbitals. For the $yz$ and $xz$ orbitals, $t_4$ and $t_2$ are the first and second neighbor hopping terms. The $\rm Os O_6$ octahedral rotation and tilting result in small mixing between the $yz$, $xz$ and $xy$ orbitals, which is represented by the first neighbor hopping terms $t_{m1}$ and $t_{m2}$. From the transformation of the hopping Hamiltonian matrix in the rotated basis, the orbital mixing hopping terms have been shown to be related to the $\rm Os O_6$ octahedral rotation and tilting angles through $t_{m1}=V_{\pi} \theta_r = t_1 \theta_r$ and $t_{m2}=V_{\pi} \theta_t /\sqrt{2} = t_1 \theta_t /\sqrt{2}$ in the small angle approximation.\cite{osmate_band_structure_2017} 

\begin{figure}
\vspace*{0mm}
\hspace*{0mm}
\psfig{figure=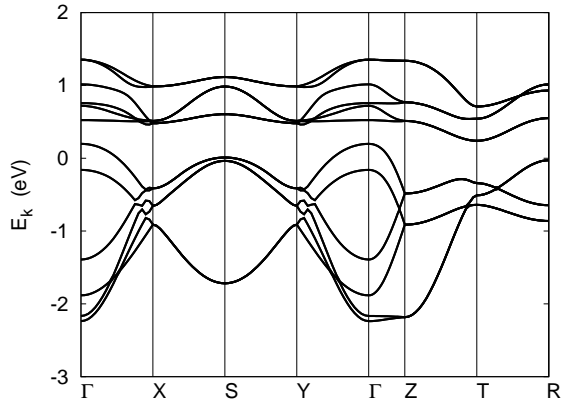,angle=0,width=55mm,angle=-90}
\caption{Electronic band structure calculated from the minimal three-orbital model for the given parameters showing the marginally insulating AFM state.} 
\label{band}
\end{figure}

The case $t_4=t_1$ and $t_{m1}=t_{m2}=0$ corresponds to the undistorted structure (cubic symmetry) with identical hopping terms for all three orbitals and no orbital mixing hopping terms. The effect of structural distortion will be approximately incorporated, within the minimal three-orbital model, through the hopping asymmetry $t_4 < t_1$, corresponding to slightly reduced bandwidth for the $yz,xz$ orbitals. We will initially neglect the orbital mixing terms $t_{m1},t_{m2}$ in order to focus on the role of the hopping asymmetry, and study their effect on magnetic anisotropy in Sec. VI.

Figure \ref{band} shows the calculated electronic band structure in the AFM state for the minimal three-orbital model with SOC ($\lambda=1.0$) and staggered field $\Delta = 1.0$ in the $z$ direction (same for all three orbitals). The hopping parameters $t_1,t_2,t_4,\epsilon_{xy} = -1.0,0.3,-0.7,0.0$. The energy scale $t_1=400$ meV corresponding to the distorted structure,\cite{osmate_band_structure_2017} which yields $\lambda = 0.4$ eV and $\Delta=0.4$ eV. In general, the strength of SOC in osmates is about 0.3 - 0.4 eV, so the SOC value taken above is consistent with this range. The separation of the energy bands into two groups of three above and three below the Fermi energy corresponds to the scenario where Hund's rule coupling dominates over spin-orbit coupling. Strong SOC significantly reduces the indirect band gap (conduction band minimum at T), yielding a marginally insulating AFM state. The electronic band structure for the minimal three-orbital model is broadly consistent with the realistic three-orbital model calculation and DFT result.\cite{osmate_band_structure_2017}  Henceforth, the values of $t_1,t_4$ (both negative) will refer to their magnitudes, and values of $t_2,\epsilon_{xy}$ will be as given above, unless specifically mentioned.

\section{SOC induced magnetic anisotropy}

The local spin-orbit coupling terms (Eq. A1) explicitly break the SU(2) spin-rotation symmetry. In accordance, a strong-coupling expansion explicitly shows the emergence of anisotropic spin interactions (Appendix A). However, when all three contributions are considered together (Eq. A2), the magnetic anisotropy is expressed only when the magnetic moments are orbitally different (Eq. A3). The hopping asymmetry $t_4 < t_1$ and the resulting magnetic moment asymmetry $m_{yz},m_{xz} > m_{xy}$ is therefore an essential requirement for the expression of magnetic anisotropy. 

In this section we will investigate the SOC-induced magnetic anisotropy and preferential ordering direction for the $(\pi,\pi,\pi)$ AFM state of the minimal three-orbital model in terms of the AFM state energy for different orientations of the staggered field. For simplicity, we will consider the same staggered field for all three orbitals. The AFM state energy (per state) was obtained by summing the HF level band energies over the occupied states.  

\begin{figure}
\vspace*{0mm}
\hspace*{0mm}
\psfig{figure=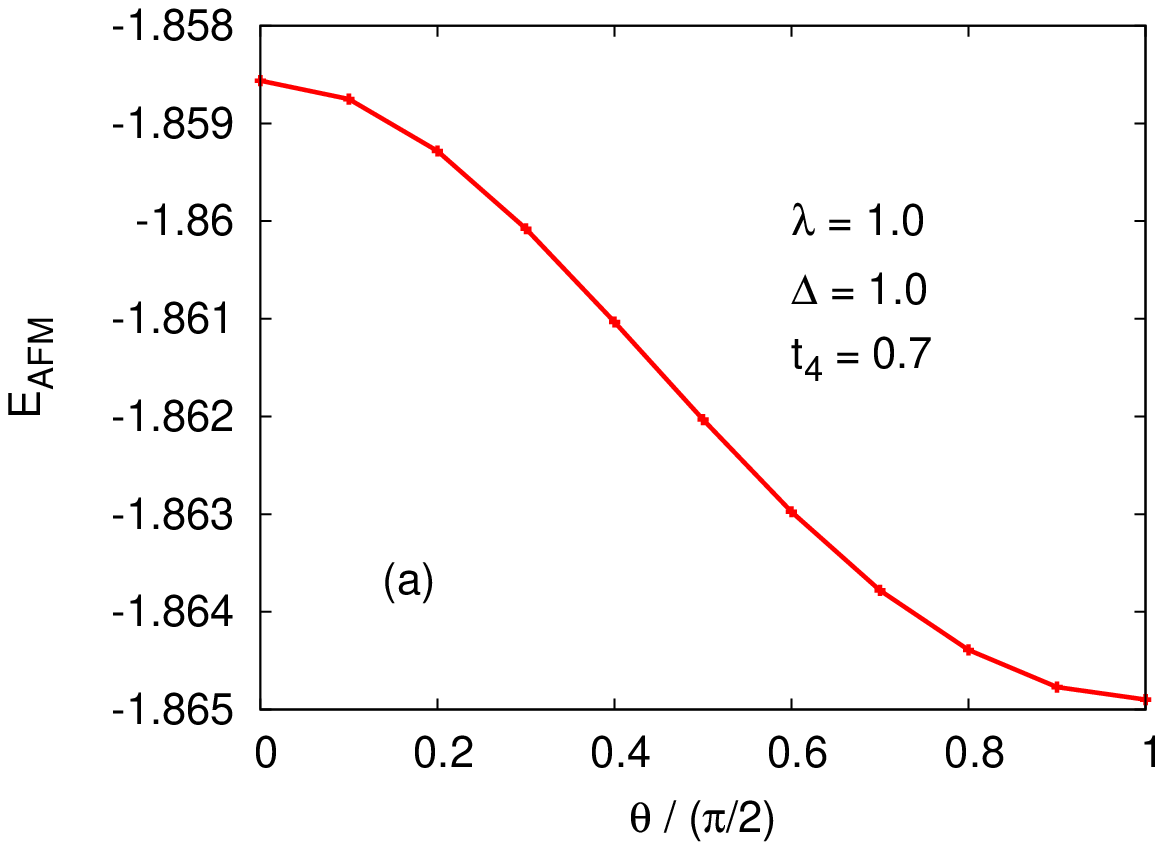,angle=0,width=80mm}
\psfig{figure=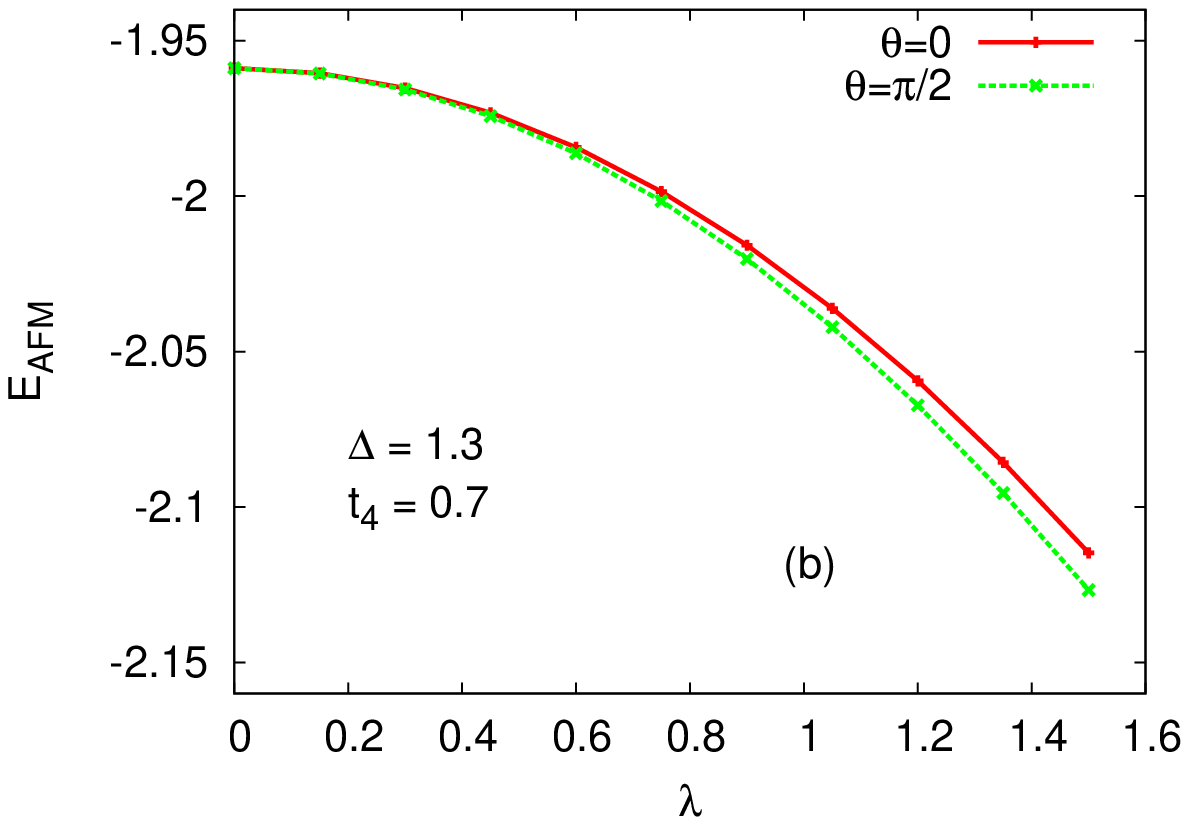,angle=0,width=80mm}
\caption{SOC induced magnetic anisotropy shown by significant dependence of AFM state energy with (a) the staggered field orientation $\theta$ from the $S_z$ axis and (b) the SOC strength for two different orientations $\theta = 0$ and $\pi/2$. The easy $x-y$ plane anisotropy and the $\cos ^2 \theta$ dependence correspond to the single-ion anisotropy term $D S_{iz} ^2$.} 
\label{gse}
\end{figure}

\begin{figure}
\vspace*{0mm}
\hspace*{0mm}
\psfig{figure=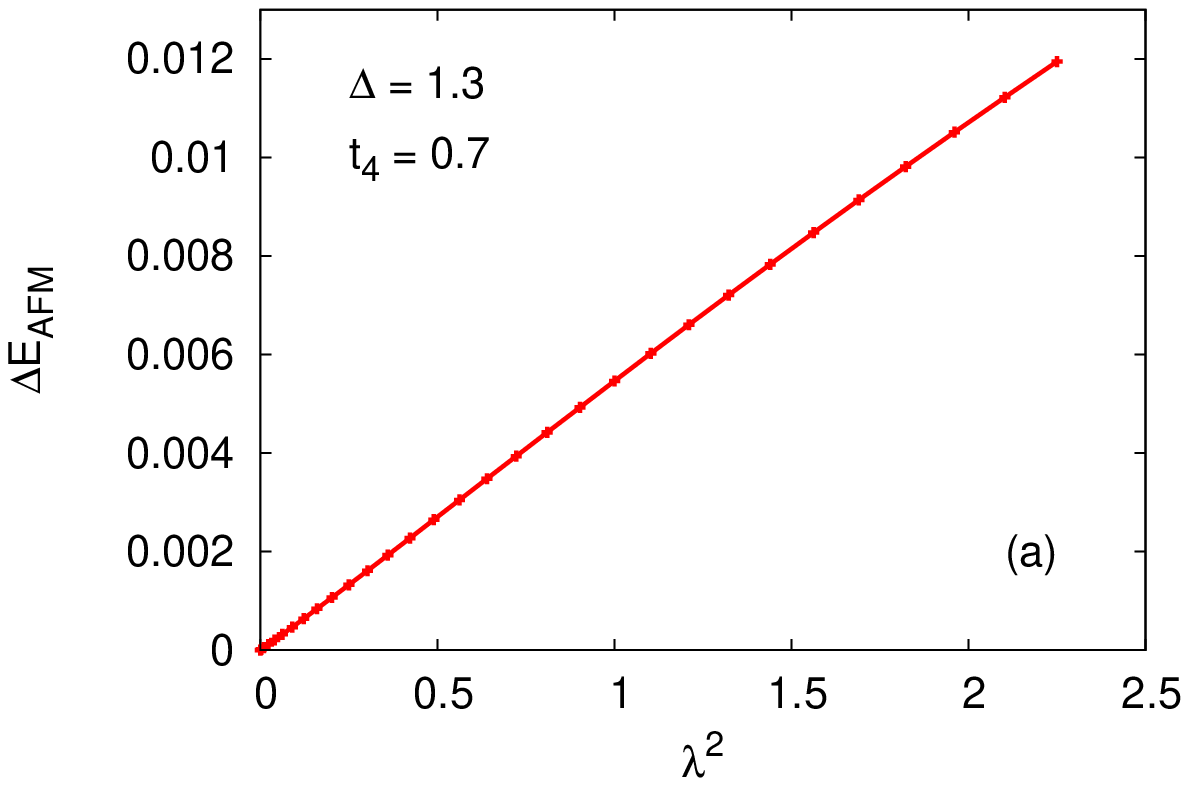,angle=0,width=80mm}
\psfig{figure=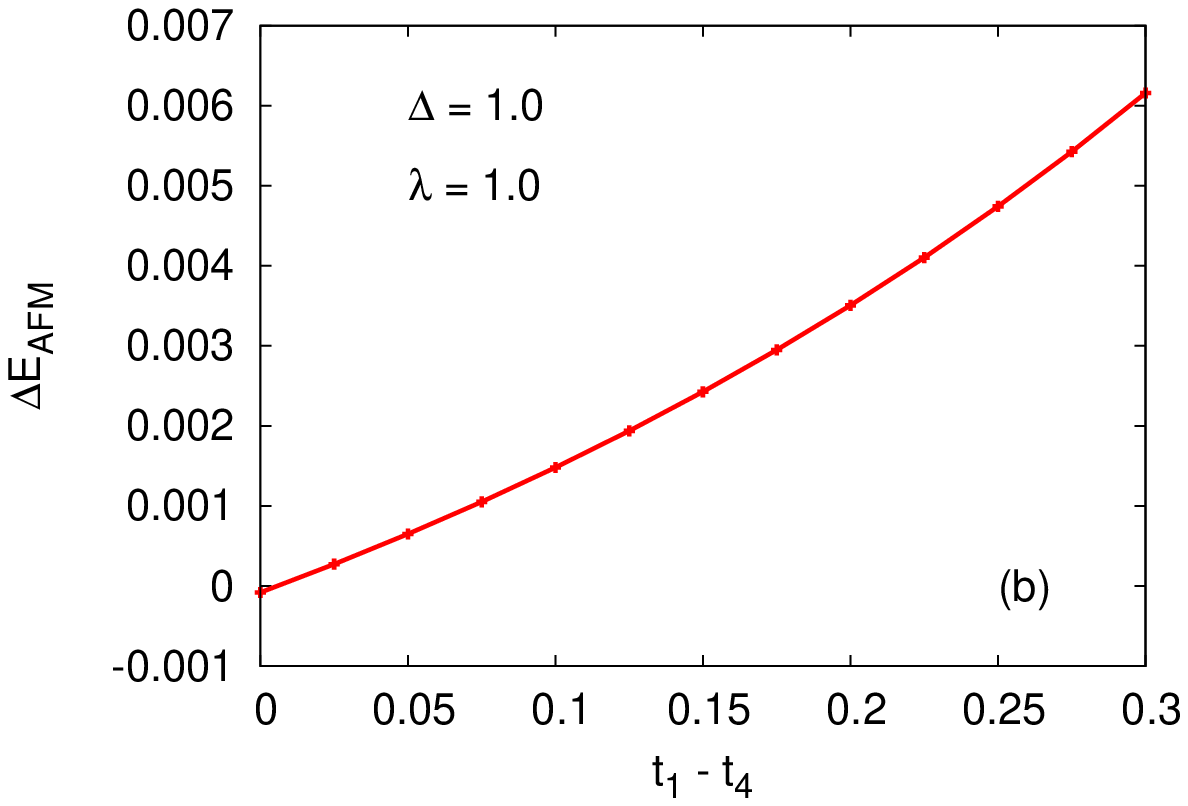,angle=0,width=80mm}
\caption{Variation of the magnetic anisotropy energy $\Delta E_{\rm AFM} = E_{\rm AFM} (z) - E_{\rm AFM} (x)$ with (a) the square of the spin-orbit coupling constant and (b) the hopping asymmetry $t_1 - t_4$ between the $xy$ and the $yz/xz$ orbitals.} 
\label{gsediff}
\end{figure}

The SOC-induced magnetic anisotropy is shown in Fig. \ref{gse}. The AFM state energy $E_{\rm AFM}$ decreases quadratically with SOC strength, the reduction being weakly dependent on the staggered field orientation. In the absence of SOC, $E_{\rm AFM}$ is independent of $\theta$. The magnetic anisotropy energy $\Delta E_{\rm AFM} = E_{\rm AFM} (z) - E_{\rm AFM} (x)$ evaluated from $E_{\rm AFM}$ for $z$ and $x$ orientations of the staggered field varies as $\lambda^2$ [Fig. \ref{gsediff}(a)], and crucially depends on the hopping asymmetry between the $xy$ and $yz/xz$ orbitals [Fig. \ref{gsediff}(b)]. All of these features can be readily understood from the SOC-induced anisotropic spin interactions being activated by the magnetic moment asymmetry resulting from the hopping asymmetry (Appendix A).  

\subsection*{Effective single-ion anisotropy}

For the hopping asymmetry $t_4 < t_1$, easy $x-y$ plane anisotropy was obtained. This corresponds to the single-ion anisotropy term $D S_{iz}^2$ in an effective spin model with $D > 0$. A small hopping asymmetry between the $yz$ and $xz$ orbitals further allows for easy axis selection within the $x-y$ plane. From the calculated MAE $\Delta E_{\rm AFM} \approx 0.006$ (per state) as in Fig. \ref{gse}, and using the energy scale $t_1 = 400$ meV, we obtain the effective single-ion anisotropy energy:
\begin{equation}
\Delta E_{\rm sia} = 3 \times 0.006 \times 400 \;{\rm meV} \approx 7 \; {\rm meV}
\label{sia}
\end{equation}
where the factor 3 corresponding to the three $t_{2g}$ orbitals per Os accounts for the conversion from average energy per state to average energy per ion. The MAE value is enhanced to about 9 meV when a positive orbital energy offset $\epsilon_{xy}$ is included (Sec. VI). Our calculated MAE is in agreement with the effective single-ion anisotropy energy $\Delta E_{\rm sia} = D S_{iz}^2$ for $D=4$ meV and $S=3/2$ as considered phenomenologically in recent spin wave calculations using a localized spin model.\cite{calder_PRB_2017} 

\begin{figure}
\vspace*{-10mm}
\hspace*{0mm}
\psfig{figure=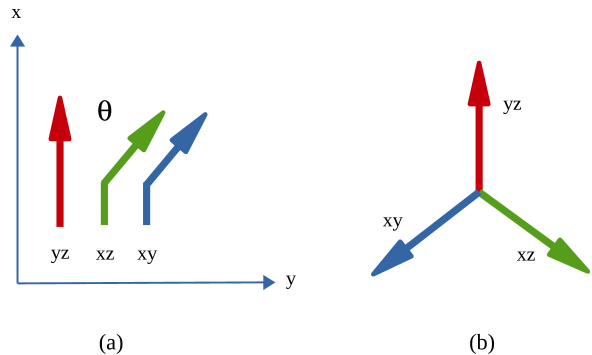,angle=0,width=100mm}
\vspace*{-60mm}
\caption{(a) Local magnetic moment orientations for the three orbitals in the canted state, and (b) the orbital 120$^\circ$ state after the transformation described in Appendix B for $\theta=\pi/3$.} 
\label{canted}
\end{figure}

\section{Intra-site magnetic frustration and orbital 120$^\circ$ state} 

\begin{figure}
\vspace*{0mm}
\hspace*{0mm}
\psfig{figure=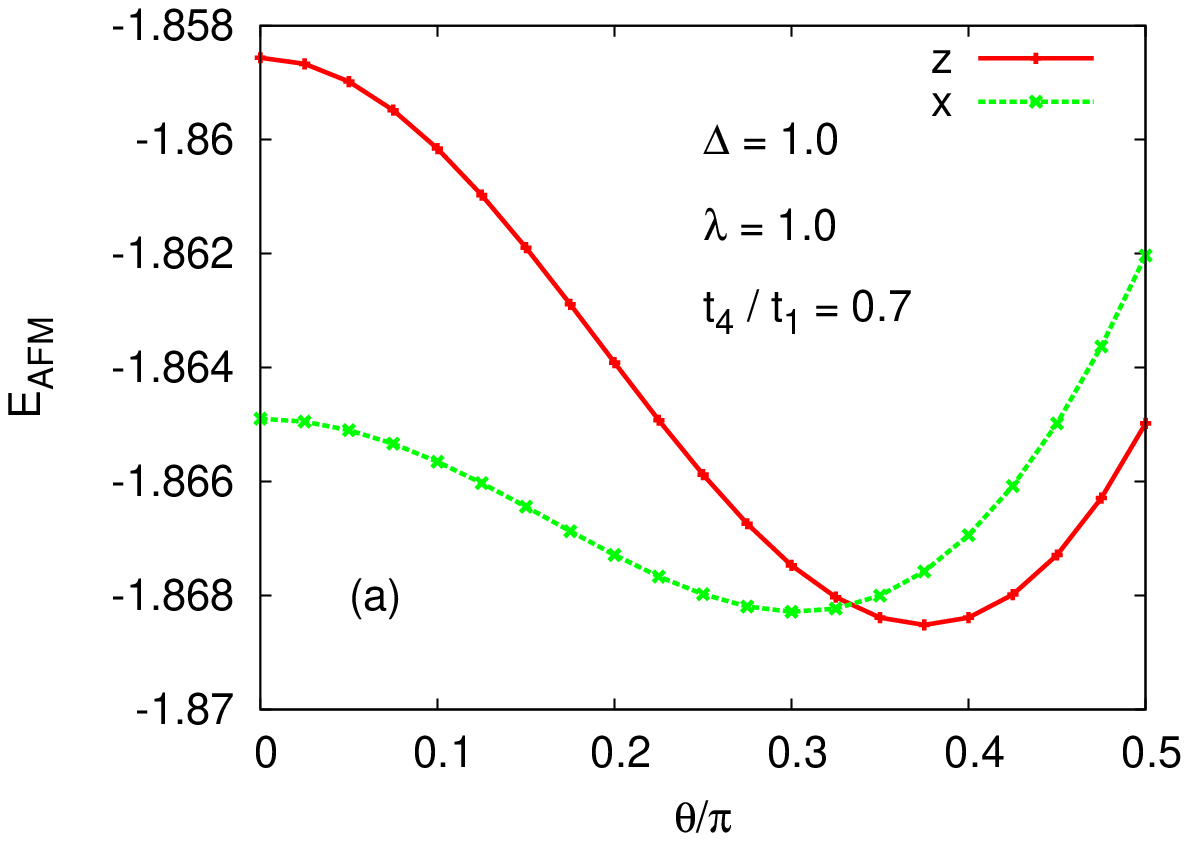,angle=0,width=80mm}
\psfig{figure=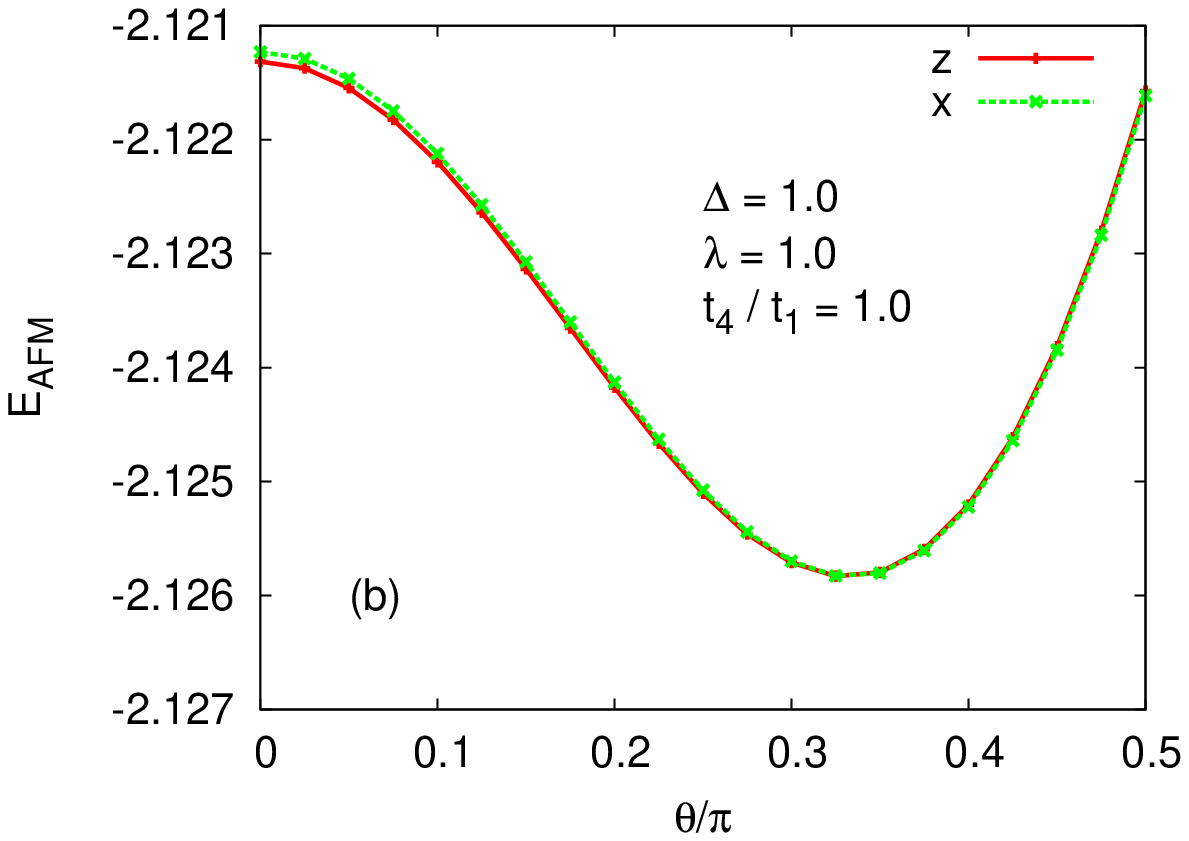,angle=0,width=80mm}
\caption{(a) Variation of the AFM state energy with angle $\theta$ in the canted state (see Fig. \ref{canted}) showing the energy minimum at $\theta \approx \pi/3$. (b) The optimal canting angle is exactly $\theta = \pi/3$ in the absence of hopping asymmetry, and the two configurations are degenerate.} 
\label{gse_vs_theta}
\end{figure}

The collinear AFM order discussed above with local magnetic moments for all three orbitals aligned parallel along some direction in the $x-y$ plane, although energetically better than ordering in the $z$ direction, is not the optimal configuration in the absence of Hund's coupling. Lower AFM state energy is obtained for the canted configuration shown in Fig. \ref{canted} where magnetic moments for two orbitals are canted with respect to the third by angle $\theta$. The ground state energies are shown in Fig. \ref{gse_vs_theta} for two configurations: (i) the $yz$ moment aligned along $x$ direction and the $xz,xy$ moments canted by angle $\theta$ in the $y$ direction and (ii) the $xy$ moment aligned along $z$ direction and the $yz,xz$ moments canted by angle $\theta$ in the $x$ direction. These two configurations are labelled $x$ and $z$, respectively. The results clearly show an energy minimum at canting angle $\theta \approx \pi/3$. The optimal canting angle is exactly $\theta = \pi/3$ in the absence of hopping asymmetry ($t_4=t_1$), and the two configurations are degenerate [Fig. \ref{gse_vs_theta}(b)].

The above proclivity towards canting of magnetic moments can be readily understood from the anisotropic spin interaction terms generated in the strong-coupling expansion (Appendix A). Assuming equal magnitudes for the magnetic moments $S_\mu$, the classical energy contribution for the canted configuration:
\begin{eqnarray}
\Delta E_{\rm SOC} (\theta) & = & \frac{4 (\lambda/2)^2 }{U} S_\mu ^2 [(-\cos \theta ) + (\cos^2 \theta - \sin^2 \theta ) + (-\cos \theta)] \nonumber \\
& = & \frac{4 (\lambda/2)^2 }{U} S_\mu ^2 [\cos 2 \theta - 2\cos \theta ]
\end{eqnarray}
corresponding to the three terms in Eq. (A2). Minimization yields $\theta = \pi/3$, with equal contribution from each of the three terms to the minimum energy $-(3/2)(\lambda^2/U)S_\mu ^2$, whereas the energy for the collinear configuration is $-(\lambda^2/U)S_\mu ^2$. 
  
Including the additional canted-state energy contribution:
\begin{equation}
\Delta E_{\rm H} = -2 J_{\rm H} S_\mu ^2 (1 + 2 \cos \theta )
\end{equation} 
from the Hund's coupling term $- 2 J_{\rm H} \sum_{\mu \neq \nu} {\bf S}_\mu . {\bf S}_\nu $, we obtain:
\begin{equation}
\Delta E_{\rm SOC} (\theta) + \Delta E_{\rm H} (\theta) = J_\lambda S_\mu ^2 [\cos 2 \theta - 2(1 + r_{\rm H}) \cos \theta - r_{\rm H}]
\end{equation}
where $J_\lambda \equiv 4(\lambda/2)^2/U$ and the ratio $r_{\rm H} = 2J_{\rm H}/J_\lambda$. Minimization of Eq. (9) now yields $\cos \theta = (1+r_{\rm H})/2$ or $\sin \theta = 0$,
and the optimal canting angle decreases from $\theta = \pi/3$ at $r_{\rm H}=0$ to $\theta = 0$ for $r_{\rm H} \ge 1$, as expected with increasing Hund's coupling. 


The significant role of Hund's coupling on magnetic anisotropy is evident from Fig. \ref{gse_vs_theta}(a). In the absence of $J_{\rm H}$, the energy minima at canting angle $\theta \approx \pi/3$ are nearly degenerate for the two configurations labelled $x$ and $z$. However, with the canting angle $\theta$ reduced to zero at sufficiently strong $J_{\rm H}$, the magnetic anisotropy is activated, favouring ordering in the $x-y$ plane as compared to the $z$ direction.  

\section{Spin wave excitations} 

Due to the presence of spin mixing terms in the Hamiltonian (Eqs. \ref{three_orb_two_sub} and \ref{gen_ord_dirn}), spin is not a good quantum number, and we therefore use the general method to investigate spin waves.\cite{tri} In the $(\pi,\pi,\pi)$ AFM ground state $|\Psi_0 \rangle$ of the three-orbital model, we consider the time-ordered transverse spin fluctuation propagator in the composite orbital-sublattice basis:
\begin{equation}
\chi^{-+}({\bf q},\omega) = \int dt \sum_i e^{i\omega(t-t')} 
e^{-i{\bf q}.({\bf r}_i -{\bf r}_j)} 
\times  \langle \Psi_0 | T [S_{i\mu} ^\alpha (t) S_{j\nu} ^\beta (t')] |\Psi_0 \rangle 
\end{equation}
involving the spin operators at lattice sites $i,j$ for orbitals $\mu,\nu$ and components $\alpha,\beta=x,y,z$. In the random phase approximation (RPA), the spin wave propagator is obtained as:
\begin{equation}
[\chi^{-+} _{\rm RPA} ({\bf q},\omega)] = \frac{[\chi^0 ({\bf q},\omega)]}{{\bf 1} - 2[U][\chi^0 ({\bf q},\omega)]}
\label{rpa}
\end{equation}
where the local interaction matrix $[U]$ in the orbital-sublattice basis is given by: $[U]_{\mu \nu} = U_\mu$ for $\mu = \nu$ (intra-orbital Hubbard term) and $[U]_{\mu \nu} = J_{\rm H}$ for $\mu \ne \nu$ (inter-orbital Hund's coupling term). The bare particle-hole propagator:
\begin{equation}
[\chi^0 ({\bf q},\omega)]_{a b} ^{\alpha \beta} = \frac{1}{4} \sum_{{\bf k}, l, m} \left [ 
\frac{ 
\langle \phi_{{\bf k},l} | \sigma^\alpha | \phi_{{\bf k-q},m} \rangle_a
\langle \phi_{{\bf k-q},m} | \sigma^\beta | \phi_{{\bf k},l} \rangle_b 
} 
{ E^+_{{\bf k-q},m} - E^-_{{\bf k},l} + \omega - i \eta }
+ \frac{
\langle \phi_{{\bf k},l} | \sigma^\alpha | \phi_{{\bf k-q},m} \rangle_a
\langle \phi_{{\bf k-q},m} | \sigma^\beta | \phi_{{\bf k},l} \rangle_b 
} 
{ E^+_{{\bf k},l} - E^-_{{\bf k-q},m} + \omega - i \eta } \right ]
\end{equation}
was evaluated in the orbital-sublattice basis by integrating out the fermions in the AFM state. Here $E_{\bf k}$ and $\phi_{\bf k}$ are the eigenvalues and eigenvectors of the Hamiltonian matrix, the indices $a,b$ in the composite orbital-sublattice basis run through 1-6, and $l,m$ indicate the eigenvalue branches. The superscripts $+$ $(-)$ refer to particle (hole) energies above (below) the Fermi energy. The spin wave energies $\omega_{\bf q}$ were obtained from the poles of Eq. \ref{rpa}.

\begin{figure}
\vspace*{0mm}
\hspace*{0mm}
\psfig{figure=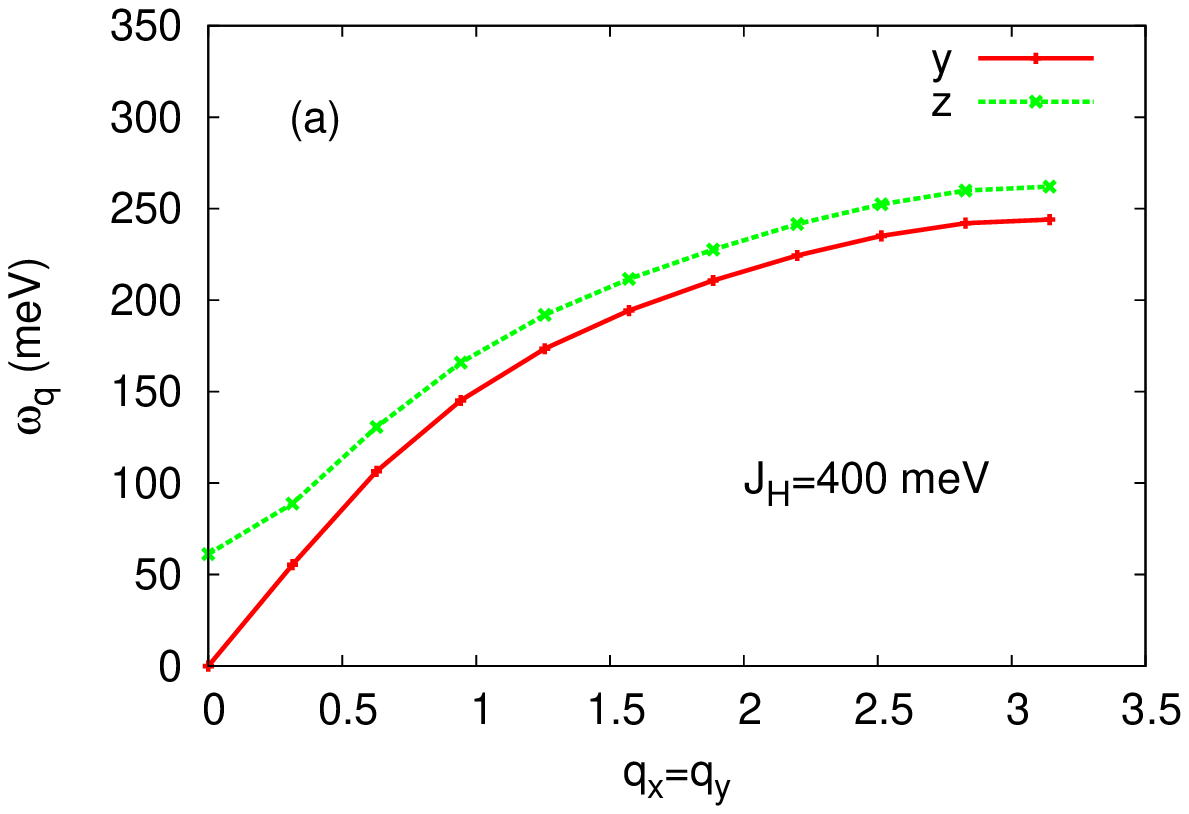,angle=0,width=80mm}
\psfig{figure=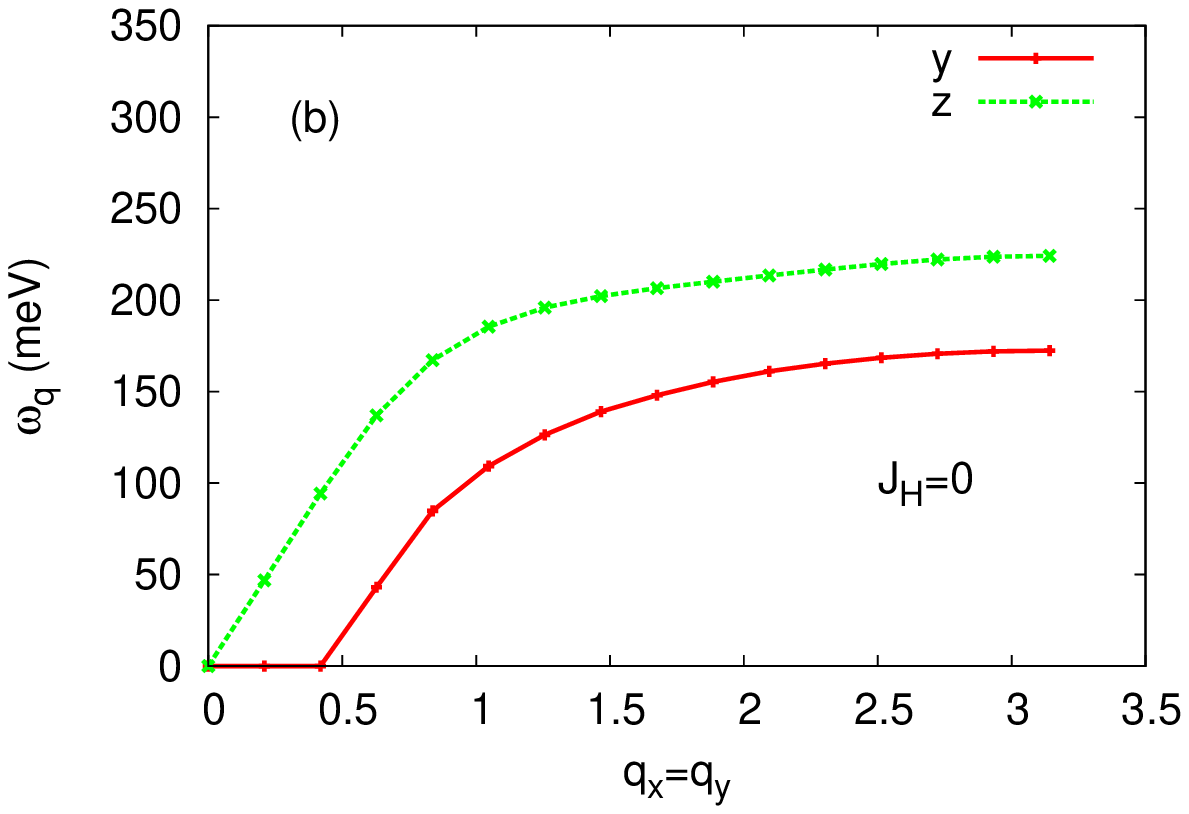,angle=0,width=80mm}
\caption{Calculated spin wave energies in the orbitally collinear AFM state ($x$ direction ordering) for the two cases: (a) with and (b) without Hund's coupling. The finite spin wave gap $\approx 60$ meV for the $z$ fluctuation mode in (a) corresponds to the finite MAE at $\theta=0$ in Fig. \ref{gse_vs_theta}(a). The instability of the $y$ fluctuation mode in (b) corresponds to the AFM state energy minimum not at $\theta=0$ but at finite canting angle in the absence of Hund's coupling.} 
\label{wq_jh}
\end{figure}

Before presenting the spin wave calculation results, we will consider the role of Hund's coupling on magnetic excitations in the AFM state based on the analysis in the previous section. In the absence of Hund's coupling, the AFM state energy was shown to have a minimum at finite canting angle $\theta \approx \pi/3$ [Fig. \ref{gse_vs_theta}(a)]. As the orbitally collinear AFM state ($\theta = 0$) with magnetic moments for all three orbitals aligned parallel (in the $x$ direction) does not correspond to the ground state configuration, spin waves should therefore yield negative energy mode representing the instability of this AFM state. Furthermore, as the optimal canting angle decreases when Hund's coupling is turned on, approaching $\theta=0$ for sufficiently strong $J_{\rm H}$, the orbitally collinear AFM state ($\theta=0$) with moments ordered in the $x$ direction now does represent the ground state, and spin waves should therefore yield gapless mode corresponding to transverse fluctuations in the $y$ direction, whereas fluctuations in the $z$ direction should become gapped. 

Calculated spin wave energies in the orbitally collinear AFM state with magnetic moments for all three orbitals oriented along the $x$ direction are shown in Fig. \ref{wq_jh}. Here the staggered field $\Delta=1.1$, $\lambda = 1$, $t_1=-1.0$, $t_2=0.3$, and $t_4=-0.7$, with the energy scale $|t_1|=400$ meV. For $J_{\rm H}=0$, the negative energy of the $y$-fluctuation mode near $q=0$ confirms the instability as expected from Fig. \ref{gse_vs_theta}(a). However, for sufficiently strong $J_{\rm H}$, when the optimal canting angle decreases to $\theta = 0$, the finite MAE accounts for the large spin wave gap $\approx 60$ meV seen in Fig. \ref{wq_jh}(a) for the out-of-plane $z$ fluctuation mode. 

\section{Effect of the orbital energy offset}

\begin{figure}
\vspace*{0mm}
\hspace*{0mm}
\psfig{figure=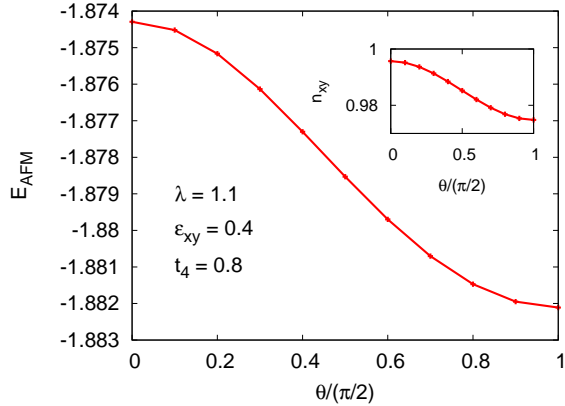,angle=0,width=80mm}
\caption{SOC induced magnetic anisotropy in presence of both hopping asymmetry $t_4 < t_1$ and orbital energy offset $\epsilon_{xy}$, as shown by the dependence of $E_{\rm AFM}$ with orientation $\theta$. Inset shows the small reduction in the $xy$ orbital density with $\theta$.} 
\label{gse_offset}
\end{figure}

The $xy$ orbital density ($n_{xy}$) is found to exhibit an intrinsic SOC-induced reduction as the staggered field orientation is rotated from $z$ direction ($\theta = 0$) to $x$ direction ($\theta = \pi/2$). This suggests that a positive energy offset $\epsilon_{xy}$ (or, equivalently, negative energy offset for $yz,xz$ orbitals, or a combination of both) should also contribute to the MAE, resulting in easy $x$-$y$ plane anisotropy. We have therefore included a small positive $\epsilon_{xy}$ (possibly arising from tetragonal distortion of the $\rm OsO_6$ octahedra) which couples with the SOC-induced reduction in $n_{xy}$, and evaluated the AFM state energy variation with $\theta$ [Fig. \ref{gse_offset}]. For the parameters shown, the MAE $\Delta E_{\rm AFM} = E_{\rm AFM} (z) - E_{\rm AFM} (x) \approx 0.0075$, which yields $\Delta E_{\rm sia} \approx 9$ meV from Eq. \ref{sia}, with roughly equal contributions from the orbital energy offset $\epsilon_{xy}$ and the hopping asymmetry $t_4 < t_1$ when considered individually. The staggered field magnitudes were taken as $\Delta_{yz}=\Delta_{xz}=1.2$ and $\Delta_{xy}=1.1$ such that $U_\mu \approx 3.5$ for all three orbitals ($J_{\rm H}=0$). 

Similar magnitude of the MAE was obtained within the three orbital model with realistic hopping parameters obtained by comparing the electronic band structure with DFT results.\cite{osmate_band_structure_2017} This calculation included the orbital mixing hopping terms ($t_{m1},t_{m2}$) given in Eq. \ref{three_orb_model}. The MAE was found to be slightly enhanced by $t_{m2}$  (tilting) and slightly suppressed by $t_{m1}$ (rotation), with essentially no net enhancement when both mixing terms were included. This approximate cancellation is also seen in our minimal three-orbital model. With increasing interaction strength $U$, the SOC-induced MAE due to both microscopic factors considered above is suppressed, highlighting the key role of weak correlation effect in the expression of large MAE.       

The reduction in $n_{xy}$ with staggered field rotation can be understood in terms of the evolution of the SOC-split energy levels with increasing exchange field $\Delta$ in the atomic limit.\cite{osmate_band_structure_2017} In the $\Delta \rightarrow 0$ limit, the $t_{2g}$ levels are split into the $J=1/2$ doublet and the $J=3/2$ quartet, and the total electron densities in the three lowest-energy $(J=3/2$) levels are: $n_{xy} = 8/6$ and $n_{yz} = n_{xz} = 5/6$.\cite{shubhajyoti_PRB_2017} Due to progressive suppression of the SOC-induced spin-orbital entanglement, the density disparity decreases with increasing $\Delta$. However, $n_{xy}$ remains greater than $n_{yz},n_{xz}$ for finite $\Delta$, even when hopping terms are included. Thus, for $z$ orientation of the staggered field, $n_{xy} > n_{yz},n_{xz}$, as indeed confirmed from the three-band model calculation. Now, rotating the staggered field from $z$ to $x$ direction is equivalent to spin space rotation by angle $\pi/2$ about the $y$ axis, under which the orbitals transform as: $xz \rightarrow xz$, $yz \rightarrow xy$, and  $xy \rightarrow yz$. The interchange of the $yz$ and $xy$ orbitals implies that $n_{xy} < n_{yz}$ for $x$ orientation of the staggered field. 

The above analysis highlights the importance of the residual $J=3/2$ character of valence band states and weak correlation in the magnetic anisotropy effect arising due to the coupling of the density change $n_\mu (z) - n_\mu (x)$ with the tetragonal distortion-induced orbital energy offset. Recent RIXS studies of the $5d^3$ systems $\rm Ca_3LiOsO_6$ and $\rm Ba_2YOsO_6$ have revealed evidence of the spin-orbit entangled $J=3/2$ character of the electronic ground state.\cite{taylor_PRL_2017}

\section{Conclusions}

Magnetic ordering, ground state energy, and magnetic excitation were investigated in the AFM state of a minimal three-orbital model at half filling with strong spin-orbit coupling. Small asymmetry in the hopping terms for the three orbitals $yz,xz,xy$ (associated with the $\rm Os O_6$ octahedral tilting and rotation), resulting in asymmetry in the magnetic moments, was shown to be an essential ingredient for the SOC-induced magnetic anisotropy and large spin wave gap observed in the weakly correlated 5d$^3$ compound $\rm Na Os O_3$ involving competition between SOC, Hund's coupling, and the staggered field, all having comparable energy scales. A novel canted AFM state was found to be stabilized by the intrasite magnetic frustration effect due to the SOC-induced anisotropic spin interactions. Restoration of the orbitally collinear AFM state by Hund's coupling was shown to be instrumental in the expression of the magnetic anisotropy and the large spin wave gap.

The residual $J=3/2$ character of the valence band states resulting from the combined SOC and electron interaction effects was found to exhibit a signature effect of reduction in the electron density $n_{xy}$ with staggered field rotation from $z$ to $x$ direction. Coupling of this density change with the orbital energy offset $\epsilon_{xy}$ was also found to contribute significantly to the magnetic anisotropy energy. The calculated magnetic anisotropy energy is similar to that obtained within the three orbital model with realistic hopping parameters determined from the electronic band structure comparison with DFT results.

\appendix 

\section{SOC-induced anisotropic spin interactions}

The spin-orbit coupling terms can be written in spin space as:
\begin{eqnarray} 
H_{\rm SO} & = & \sum_i \left [ \begin{pmatrix} \psi_{yz \uparrow}^\dagger & \psi_{yz \downarrow}^\dagger \end{pmatrix}
\begin{pmatrix} i \sigma_z \lambda /2 \end{pmatrix} 
\begin{pmatrix} \psi_{xz \uparrow} \\ \psi_{xz \downarrow} \end{pmatrix}
+ \begin{pmatrix} \psi_{xz \uparrow}^\dagger & \psi_{xz \downarrow}^\dagger \end{pmatrix}
\begin{pmatrix} i \sigma_x \lambda /2 \end{pmatrix} 
\begin{pmatrix} \psi_{xy \uparrow} \\ \psi_{xy \downarrow} \end{pmatrix} \right . \nonumber \\
& + & \left . \begin{pmatrix} \psi_{xy \uparrow}^\dagger & \psi_{xy \downarrow}^\dagger \end{pmatrix}
\begin{pmatrix} i \sigma_y \lambda /2 \end{pmatrix} 
\begin{pmatrix} \psi_{yz \uparrow} \\ \psi_{yz \downarrow} \end{pmatrix} \right ] 
\end{eqnarray}
which explicitly shows the SU(2) spin-rotation symmetry breaking. Here we discuss the resulting magnetic anisotropy and preferential magnetic ordering direction. For this purpose, we perform a strong-coupling expansion as for the SOC-induced spin-dependent hopping terms of the form $i${\boldmath $\sigma . t'_{ij}$}, which yield the Kitaev type anisotropic spin interactions.\cite{hc_arxiv_2017}

As the three orbital ``hopping" terms are of similar form as spin-dependent hopping, carrying out the strong-coupling expansion to second order in $\lambda$, we obtain similar anisotropic spin interactions:
\begin{eqnarray}
H^{(2)}_{\rm eff} (i) = \frac{4 (\lambda/2)^2 }{U} & \bigg ( & \left [ S_{yz}^z S_{xz}^z - (S_{yz}^x S_{xz}^x  + S_{yz}^y S_{xz}^y) - \bar{n}_{yz} \bar{n}_{xz} \right ] \nonumber \\
& + & \left [ S_{xz}^x S_{xy}^x - (S_{xz}^y S_{xy}^y  + S_{xz}^z S_{xy}^z) - \bar{n}_{xz} \bar{n}_{xy} \right ] \nonumber \\
& + & \left [ S_{xy}^y S_{yz}^y - (S_{xy}^z S_{yz}^z  + S_{xy}^x S_{yz}^x) - \bar{n}_{xy} \bar{n}_{yz}  \right ] \bigg )
\label{h_eff}
\end{eqnarray}
which are, it should be emphasized, local (intra-site) interactions between the magnetic moments for the three orbitals at site $i$. Assuming the local magnetic moments ${\bf S}_\mu$ to be independent of the orbital index $\mu$, and similarly for the spin-averaged electron densities $\bar{n}_\mu$, we obtain:
\begin{equation}
H^{(2)}_{\rm eff} (i) = - \frac{4 (\lambda/2)^2 }{U} [ {\bf S}.{\bf S} + 3 \bar{n}^2 ]
\end{equation}
This accounts for the quadratic reduction of the AFM state energy with the SOC strength $\lambda$, as seen in Fig. \ref{gse}. The weak orbital dependence of the magnetic moments accounts for the small variation in the AFM state energy with staggered field orientation, which is the source of the magnetic anisotropy. If the magnetic moment $S_{xy}$ for the $xy$ orbital is slightly smaller than for the $xz,yz$ orbitals, and assuming parallel alignment of the magnetic moments for the three orbitals due to Hund's coupling, the term in the first line of Eq. \ref{h_eff} dominates, resulting in preferred  ordering in the $x-y$ plane. Within an equivalent spin model, this would correspond to the single ion anisotropy term $D S_{iz}^2$ with positive $D$.  

The preferred magnetic ordering direction within the $x-y$ plane can be further selected if the degeneracy between the $yz$ and $xz$ magnetic moments is lifted. Considering only the $x,y$ components of the magnetic moments ${\bf S}_\mu$ in Eq. \ref{h_eff}, we have: 
\begin{equation}
H^{(2)}_{\rm eff} (i) = \frac{4 (\lambda/2)^2 }{U} \bigg ( -{\bf S}_{yz} . {\bf S}_{xz} 
+ S_{xy}^x (S_{xz}^x - S_{yz}^x) + S_{xy}^y (S_{yz}^y - S_{xz}^y) \bigg )
\label{h_eff_xy}
\end{equation}
which clearly shows $x \; (y)$ to be the preferred ordering direction if the moment $S_{yz}$ is greater (less) than the moment $S_{xz}$. Without sufficiently strong Hund's coupling, the orbitally collinear AFM state with the magnetic moments ${\bf S}_\mu$ for all three orbitals aligned parallel does not correspond to the lowest-energy state due to the intra-site magnetic frustration effect, as discussed below. 

\section{Spin transformation, magnetic frustration, and orbital 120$^\circ$ state}
Under the transformation: 
\begin{eqnarray}
{\bf S}_{yz} \rightarrow {\bf S}_{yz}' & = & (S_{yz}^x, -S_{yz}^y, -S_{yz}^z) \nonumber \\
{\bf S}_{xz} \rightarrow {\bf S}_{xz}' & = & (-S_{xz}^x, S_{xz}^y, -S_{xz}^z) \nonumber \\
{\bf S}_{xy} \rightarrow {\bf S}_{xy}' & = & (-S_{xy}^x, -S_{xy}^y, S_{xz}^z)
\label{transf}
\end{eqnarray}
where two spin components are reversed for each orbital in cyclic fashion, the effective spin interaction Hamiltonian (Eq. \ref{h_eff}) transforms to the isotropic form:
\begin{equation}
H^{(2)} _{\rm eff}  (i) = \frac{4 (\lambda/2)^2 }{U} \sum_{\mu \ne \nu} 
({\bf S' _\mu}.{\bf S' _\nu} - n_\mu n_\nu)
\label{h_iso}
\end{equation}
which highlights the SOC-induced magnetic frustration between the three local magnetic moments ${\bf S'}_\mu$. In analogy with the 120$^\circ$ state of the geometrically frustrated triangular lattice AFM, the orbital canted state shown in Fig. \ref{canted}(a) corresponds, for $\theta=\pi/3$, to an orbital 120$^\circ$ state in which the transformed magnetic moments ${\bf S}_{yz}',\; {\bf S}_{xz}',\; {\bf S}_{xy}'$ are oriented at 120$^\circ$ with respect to each other, as shown in Fig. \ref{canted}(b). 

This intra-site magnetic frustration and canting tendency of the local magnetic moments persists even when hopping is turned on, as is evident from Fig. \ref{gse_vs_theta}, showing the energy minimum at canting angle $\theta \approx \pi/3$ in the band AFM state.

\section*{Acknowledgement}
We thank the U.S. Department of Energy, Office of Basic Energy Sciences, Division of Materials Sciences and Engineering (Grant No. DE-FG02-00ER45818) for partial financial support. AS acknowledges sponsorship grant from the Alexander von Humboldt Foundation for a research stay at IFW Dresden. 


\end{document}